\documentclass[twocolumn]{aastex631} 

\usepackage[english]{babel}
\usepackage{amsmath}
\usepackage{amssymb}
\usepackage{graphicx}
\usepackage{subfigure}
\usepackage[normalem]{ulem}
\usepackage{enumerate}
\usepackage{booktabs}

\usepackage{verbatim}

\usepackage{color}
\usepackage{multirow}
\usepackage{mathtools}
\usepackage{epstopdf}
\usepackage{tabularx}

\usepackage{url}
\usepackage{xcolor}

\def\apjl{ApJL}
\def\apjs{ApJS}
\def\aap{A\&A}

\def\be{\begin{equation}} 
\def\ee{\end{equation}} 
\def\ba{\begin{eqnarray}} 
\def\ea{\end{eqnarray}}

\def\mpc{\,{\rm {Mpc}}}

\def\msun{{\Msun}}

\def\gsim{\lower.5ex\hbox{\gtsima}} 
\def\lsim{\lower.5ex\hbox{\ltsima}} \def\gtsima{$\; \buildrel > \over 
\sim \;$} \def\ltsima{$\; \buildrel < \over \sim \;$} \def\prosima{$\; 
\buildrel \propto \over \sim \;$} \def\gsim{\lower.5ex\hbox{\gtsima}} 
\def\lsim{\lower.5ex\hbox{\ltsima}} 
\def\simgt{\lower.5ex\hbox{\gtsima}} 
\def\simlt{\lower.5ex\hbox{\ltsima}} 
\def\simpr{\lower.5ex\hbox{\prosima}}   
  
 \def\gtsima{$\; \buildrel > \over \sim \;$} 
\def\ltsima{$\; \buildrel < \over \sim \;$} 
\def\gsim{\lower.5ex\hbox{\gtsima}} 
\def\lsim{\lower.5ex\hbox{\ltsima}} 
\def\simgt{\lower.5ex\hbox{\gtsima}} 
\def\simlt{\lower.5ex\hbox{\ltsima}} 
\def\simpr{\lower.5ex\hbox{\prosima}}

\def\msun{\,{\rm \Msun}}

\def\E3{{\cal E}_{\rm g}^{III}}

\def\Msun{\rm M_\odot}

\def\r12{r_{1/2}} 
\def\x12{x_{1/2}} 
\def\v12{v_{1/2}}

\newcommand\code[1]{\textsc{\MakeLowercase{#1}}}
\newcommand{\quotes}[1]{``#1''}

\def\nh2{n_{\rm H2}}
\def\fh2{f_{\rm H2}}

\def\arcsec{^{\prime\prime}}
\def\angstrom{\textrm{A\kern -1.3ex\raisebox{0.6ex}{$^\circ$}}}
\def\Myr{\rm Myr}

\def\msun{{\rm M}_{\odot}}

\def\msunyr{\msun\,{\rm yr}^{-1}}

\def\kpc{{\rm kpc}}

\def\highz{$\mbox{high-}z$~}

\def\sfrratio{\rm SFR_{3}/SFR_{50}}

\makeatletter
\def\@hex@@Hex#1%
 {\if a#1A\else \if b#1B\else \if c#1C\else \if d#1D\else
  \if e#1E\else \if f#1F\else #1\fi\fi\fi\fi\fi\fi \@hex@Hex}
\makeatother

\definecolor{apcolor}{HTML}{b3003b}
\definecolor{cbcolor}{HTML}{ff0f00}
\definecolor{afcolor}{HTML}{b3443c}
\definecolor{vgcolor}{HTML}{8F00FF}
\definecolor{tbdcolor}{HTML}{E8A95E}
\definecolor{stefcolor}{HTML}{0047ab}

\graphicspath{{plots_png/},{plots/}}

\shorttitle{high-$z$ temporarily quiescent galaxies}
\shortauthors{Gelli et al.}

\begin{document}

\title{Temporarily quiescent galaxies at cosmic dawn: probing bursty star formation}

\correspondingauthor{Viola Gelli}
\email{viola.gelli@nbi.ku.dk}
\author[0000-0001-5487-0392]{Viola Gelli}
\affiliation{Cosmic Dawn Center (DAWN)}
\affiliation{Niels Bohr Institute, University of Copenhagen, Jagtvej 128, 2200 København N, Denmark}
\author[0000-0002-7129-5761]{Andrea Pallottini}
\affiliation{Scuola Normale Superiore, Piazza dei Cavalieri 7, I-56126 Pisa, Italy}
\author[0000-0001-7298-2478]{Stefania Salvadori}
\affiliation{Dipartimento di Fisica e Astronomia, Universit\'{a} degli Studi di Firenze, via G. Sansone 1, 50019, Sesto Fiorentino, Italy}
\affiliation{INAF/Osservatorio Astrofisico di Arcetri, Largo E. Fermi 5, I-50125, Firenze, Italy}
\author[0000-0002-9400-7312]{Andrea Ferrara}
\affiliation{Scuola Normale Superiore, Piazza dei Cavalieri 7, I-56126 Pisa, Italy}
\author[0000-0002-3407-1785]{Charlotte Mason}
\affiliation{Cosmic Dawn Center (DAWN)}
\affiliation{Niels Bohr Institute, University of Copenhagen, Jagtvej 128, 2200 København N, Denmark}
\author[0000-0002-6719-380X]{Stefano Carniani}
\affiliation{Scuola Normale Superiore, Piazza dei Cavalieri 7, I-56126 Pisa, Italy}
\author[0000-0002-9122-1700]{Michele Ginolfi}
\affiliation{Dipartimento di Fisica e Astronomia, Universit\'{a} degli Studi di Firenze, via G. Sansone 1, 50019, Sesto Fiorentino, Italy}
\affiliation{INAF/Osservatorio Astrofisico di Arcetri, Largo E. Fermi 5, I-50125, Firenze, Italy}

\begin{abstract}
The bursty, time-variable nature of star formation in the first billion years, as revealed by JWST, drives phases of \textit{temporary quiescence} in low-mass galaxies that quench after starbursts. These galaxies provide unique probes of the burstiness of early star formation and its underlying physical processes.
Using the \code{SERRA} cosmological zoom-in simulations, we analyze over 200  galaxies with $M_\star<10^{9.5}\msun$ at $z\sim 6-8$, finding that most experience quiescent phases driven by stellar feedback, with minimal influence from environmental effects. 
The fraction of temporarily quiescent galaxies increases with decreasing mass and luminosity, representing the dominant population at $M_\star<10^8\msun$ and \mbox{$M_{UV}>-17$}.
By forward modeling their spectral energy distributions, we show that they are faint ($\langle M_{UV}\rangle = -15.6$ for $M_\star=10^{8}\msun$), have strong Balmer breaks ($> 0.5$) and no emission lines. Comparing our predicted fractions with JWST results, we find similar luminosity-dependent trends; however, the observed fractions of temporarily quiescent galaxies at $M_{UV}\sim-20$ to $-19$ are higher, suggesting that stronger feedback or additional mechanisms beyond supernovae may be at play.
We propose searching for F200W drop-outs and satellites in the proximity ($<5 \arcsec$) of massive ($>10^{10}\msun$) galaxies as effective strategies to uncover the hidden majority of faint ($M_{UV}>-17$), temporarily quiescent systems, crucial for constraining early feedback processes in low-mass galaxies.
\end{abstract}


\section{Introduction}

With its unprecedented depth and sensitivity, the James Webb Space Telescope (JWST) is enabling us to study the star formation histories (SFH) of galaxies in the first billion years of the Universe in remarkable detail. The way galaxies form their stars over time is determined by the complex interplay of gas cooling, accretion, feedback processes, and environmental factors.
Low-mass galaxies, with their shallow gravitational wells, are especially sensitive to these effects, leading to stochastic, or “bursty", SFHs characterized by strong fluctuations over short timescales \citep[e.g.][]{Furlanetto22, Iyer2024}.
During the first billion years since the Big Bang, the Universe is largely dominated by such low-mass galaxies, and bursty star formation is predicted to be common, as shown by several cosmological simulations \citep[e.g.,][]{Ma18, Sun23a, Pallottini2022, Bhagwat24}.

Constraining the level of \quotes{burstiness} in high-$z$ galaxies is important for several reasons. First, understanding the time-variability of the SFHs is key for unveiling the physics regulating it, as different feedback and accretion processes act on different timescales \citep[e.g.][]{Iyer2020, Tacchella20}.
Second, bursty star formation can have significant implications on multiple observables: the UV luminosity functions \citep{Mason23, Mirocha23, Pallottini23, Shen23, Sun23b, Gelli24b, Kravtsov24}, the ionizing photon budget \citep{Nikolic24}, the mass-metallicity relation \citep{Pallottini24}, and even the potential detectability of Pop III galaxies \citep{Katz23}.

JWST results are indeed providing increasing evidence of bursty star formation in the Epoch of Reionization (EoR). Galaxies at fixed stellar mass or luminosity exhibit large scatter in the distributions of properties, such as the star formation rate \citep[SFR, e.g.,][]{Ciesla23, Cole23}, the gas metallicity \citep[e.g.,][]{heintz:2023, morishita:2024}, stellar ages \citep[e.g.][]{Whitler2023}, emission line strengths \citep[e.g.][]{Navarro-Carrera24}, and ionizing photon production efficiencies \citep[e.g.,][]{Endsley23b, Prieto-Lyon23, Begley24}.

A key implication of bursty star formation is the possibility that galaxies may be quenched following strong starbursts, resulting in a population of \textit{temporarily quiescent} high-$z$ galaxies\footnote{To describe this particular class of galaxies, different definitions have been adopted: \quotes{mini-quenched}, \quotes{smoldering}, \quotes{napping}, \quotes{lulling}, etc. In this paper, we refer to them as \quotes{temporarily quiescent} galaxies.} \citep[e.g.,][]{Gelli20, Faisst24}. JWST has indeed also provided the very first detections of low-mass post-starburst quiescent galaxies in the EoR \citep{Strait23, Looser23, Baker25}.
The mechanisms driving this quenching are still debated: supernova (SN) feedback alone seems insufficient to abruptly suppress star formation in the short timescales observed \citep{Gelli23, Gelli24a, Dome24}, and environment may also play a role in some cases \citep[e.g.,][]{Asada24}.

Reconstructing and interpreting the evolution of individual galaxies is challenging, as recent starbursts outshine older stellar populations \citep[e.g.,][]{GimenezArteaga24, Narayanan24} and a lack of emission lines can be confused with high escape fractions \citep[e.g.][]{Topping22}. To overcome these issues, we can use a statistical approach. Exploiting large samples of high-$z$ galaxies we can probe systems that are experiencing different stages of their star formation. By looking at these galaxies as an \quotes{ensemble} and quantifying the fraction of quiescent vs star-bursting systems we can gain insight on early galaxy formation. 

In particular, the fraction of galaxies undergoing temporary quiescence is directly tied to the overall duration of these phases, encapsulating key information about the duty cycles and timescales of SFH variability.
Observations of large samples of this new class of temporarily quiescent galaxies uncovered by JWST offer a unique way to probe bursty star formation and high-$z$ quenching mechanisms. 

Recently, several efforts have been made toward the identification of galaxies experiencing temporarily quiescent phases of star formation \citep{Trussler24, Endsley24, Looser23b}, finding that they constitute a considerable fraction of the high-$z$ low-mass and low-luminosity galaxy population.
However, the physical interpretation of these results and comparison with theoretical models and simulations is complex and challenging, as observational samples are not complete in stellar mass and naturally biased toward the brighter and more luminous starburst phases of star formation \citep[e.g.,][]{Sun23a, Mason23}.

JWST is however finally enabling us to explore a broader range of SFRs of galaxies at $z>6$, capturing not only the starbursts but also the fainter phases of the galaxies' highly time-variable SFHs.

In this paper, we use results from the \code{serra} cosmological simulations \citep{Pallottini2022} combined with accurate forward modeling of galaxy spectra \citep{Gelli21} to predict the expected abundances of observable temporarily quiescent galaxies at high-$z$. By comparing these predictions with recent JWST results, we aim to constrain the burstiness of galaxies in the EoR and infer the physical processes shaping it.

\section{Methods}\label{sec:methods}

\subsection{SERRA cosmological simulations}

\code{SERRA} \citep{Pallottini2022} is a suite of cosmological zoom-in simulations that follow the evolution of typical Lyman break galaxies ($M_\star \simeq 10^{10} \Msun$) and their environments in the EoR.
\code{Music} \citep{HahnAbel11} is adopted to generate initial conditions at $z=100$ in a cosmological volume of $(20\mpc /h)^3$ by assuming a \citet{Planck14} cosmology\footnote{Throughout the paper we assume a $\Lambda$CDM model with vacuum, matter, and baryon densities in units of the critical density $\Omega_{\Lambda}= 0.692$, $\Omega_{m}= 0.308$, $\Omega_{b}= 0.0481$, Hubble constant $\rm H_0=67.8\, {\rm km}\,{\rm s}^{-1}\,{\rm Mpc}^{-1}$, spectral index $n=0.967$, and $\sigma_{8}=0.826$.}.
\code{RAMSES} \citep{Teyssier02} is used to follow the evolution of dark matter, stars, and gas by adopting baryon mass resolution of $1.2 \times 10^4 \Msun$ and spatial resolution of $\simeq 20~\,\rm pc$ in the zoom-in regions, i.e. about the mass and size of molecular clouds.
\code{SERRA} allows for on-the-fly radiative transfer via \code{RAMSES-RT} \citep{Rosdahl2013} that is coupled \citep{Pallottini19} with the photons-gas interactions handled with \code{KROME} \citep{krome}, by solving for the non-equilibrium chemical network up to molecular hydrogen (H$_2$) formation \citep{Pallottini17}.

Stars form following a \cite{schmidt:1959}-\cite{Kennicutt98} relation depending on the H$_2$ density, and assuming a \cite{Kroupa01} initial mass function.
Stellar feedback modeling includes SNe explosions, winds, and an approximate treatment of radiation pressure \citep{Pallottini16}.
\code{STARBURST99} \citep{starburst99} is used to account for the mechanical energy injection (SN and winds), photon inputs, and chemical yields, depending on the stellar age and metallicity, that are taken from \code{PADOVA} stellar tracks \citep{Bertelli94}, covering a metallicity range of $Z_\star / \rm Z_\odot = 0.02-1.0$. 
As a $1.2 \times 10^4 \Msun$ resolution does not allow to follow the evolution of the first stars and mini-halos, their effect is accounted for by setting the initial gas metallicity to a floor value of $Z_{\rm floor}=10^{-3} Z_\odot$ \citep{Wise12, Pallottini14}. 

In this work, we analyze multiple \code{SERRA} galaxies taken in the $6<z<8$ redshift range.
To quantify the importance of environmental effects on the evolution of low-mass galaxies, we select and analyze both \textit{central} galaxies and \textit{satellites} galaxies orbiting more massive systems.
However, the typical clustering algorithm for galaxy identification \citep[e.g.][]{knollmann2009, rockstar} is calibrated for large cosmological simulations and can experience problems in pinpointing the satellites located in the highest density regions, particularly in the case of zoom-in simulations.
Thus, first we use \code{ROCKSTAR} \citep{rockstar} to select galaxies $M_\star < 10^{10.5} \Msun$ and then the low-mass satellites sample is extracted using a stellar-density based method \citep[][]{Gelli20}, checking \textit{a~posteriori} for the presence of a dark matter (sub-)halo.

The sample used in this paper consists of 78 central and 131 satellite galaxies at $6<z<8$ with stellar masses in the range $M_\star = 10^7-10^{9.5} \Msun$.

\begin{figure*}
\centering
\includegraphics[width=\textwidth]{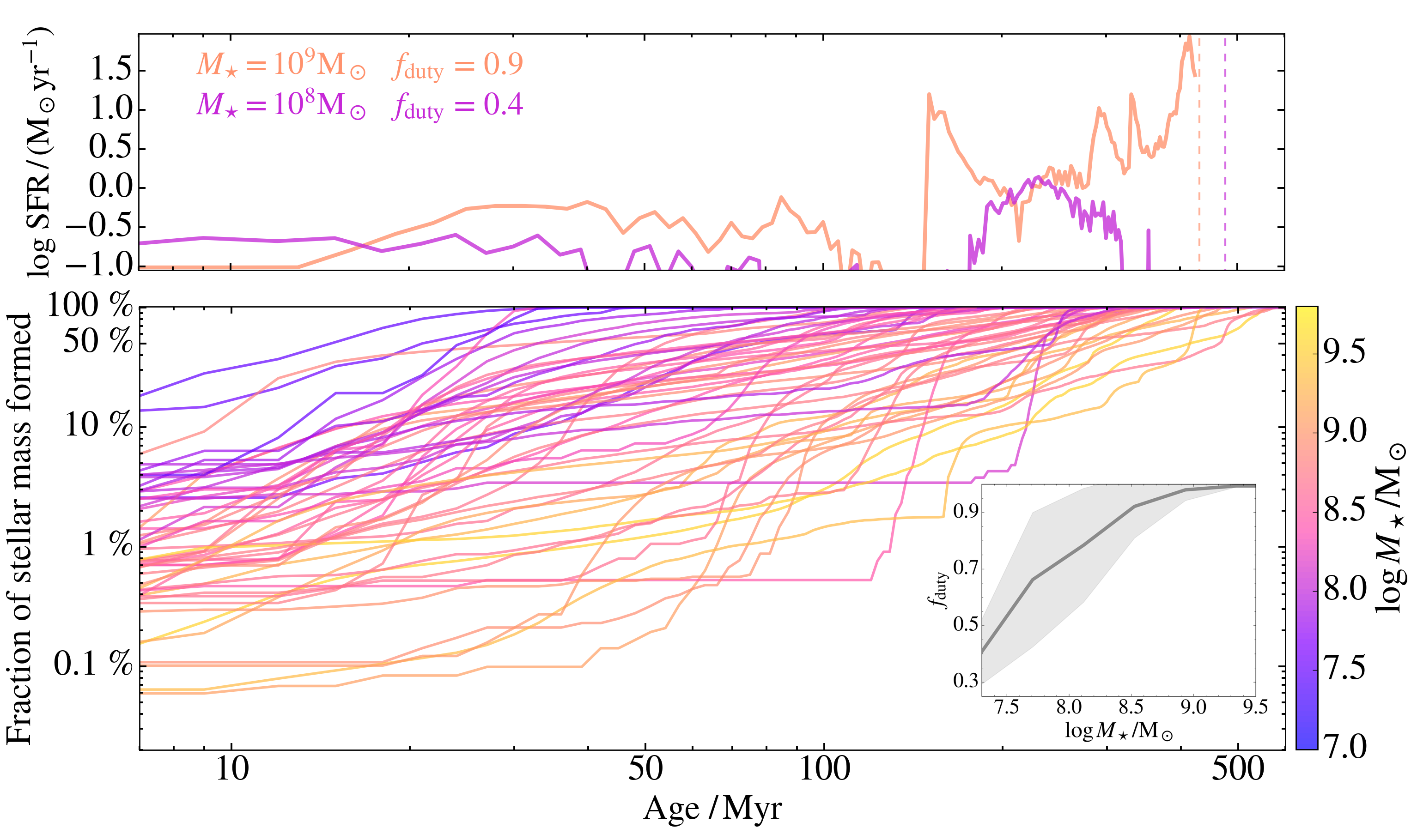}
\caption{
The top panel shows the SFH of two typical galaxies, one with high mass and high duty cycle, the other with low mass and low duty cycle and undergoing a quiescent phase at the end of the simulation ($z\sim6$, indicated by the dashed vertical line).
The bottom panel shows the fraction of stellar mass formed with respect to the final stellar mass ($M_\star$) of each galaxy as a function of the time elapsed from its first star formation event.
The colors indicate the final stellar mass of the galaxies.
Most galaxies experience periods of quiescence during their early evolutionary stages.
The inset shows the mean trend of the duty cycle, i.e. the fraction of time spent in activity, that increases as a function of the stellar mass.
\label{fig:cummass}
}
\end{figure*}

\subsection{Spectral energy distribution modeling}\label{sec:sed_modelling}

To produce synthetic spectral energy distributions (SEDs) of the simulated galaxies, firstly we model their stellar continuum emission using \code{STARBURST99} \citep{starburst99}, consistent with the simulation prescriptions.

\code{CLOUDY} \citep{Ferland17} is used to compute nebular line emission from the galaxies' interstellar medium, taking into account gas density, metallicity, turbulent structure, and radiation field \citep{vallini:2018, Pallottini19}.
We compute the main lines typically contributing to the rest-frame UV-optical spectrum (i.e. hydrogen $\rm H\alpha$, $\rm H\beta$, $\rm H\gamma$, oxygen [OII]$\lambda\lambda$3726,3729 and [OII]$\lambda\lambda$4959,5007, and carbon (CIII]1909) by summing cell-by-cell luminosities from \code{CLOUDY} and assuming a Gaussian profile with a width determined by the thermal and turbulent motions \citep[see][]{kohandel:2020}.

Finally, we take into account the presence of dust in the galaxies, which can attenuate the intrinsic spectrum \citep[i.e. see][]{Gelli21}. Specifically, we adopt a dust-to-metal ratio of $f_d = 0.08$ and assume MW-like dust \citep{Behrens18}: we use the extinction curve from \citealt{Weingartner01} to attenuate the synthetic intrinsic SED of each galaxy\footnote{As a test of the method, the continuum spectra of galaxies at $z=7.7$ in the present sample have been compared with the SEDs resulting from 3D continuum radiative transfer post-processing done via \code{SKIRT} \citep{skirt15} with the setup adopted in \citet{Pallottini2022}. The SEDs obtained with the two methods are consistent.}.
Note that the assumptions for dust differ from \citet{Gelli21}, where an SMC-like extinction curve and $f_d = 0.3$ is adopted. This change of assumptions is driven by more recent ALMA measurements of high-$z$ galaxies \citep{bouwens:2022}, that support a MW-like dust \citep{Ferrara22} and lower dust-to-metal ratios \citep[][]{Laporte17a, Behrens18}. However, the main results do not change significantly with respect to the previous assumptions when considering these changes.

\begin{figure*}[t!]
\centering
\includegraphics[width=\textwidth]{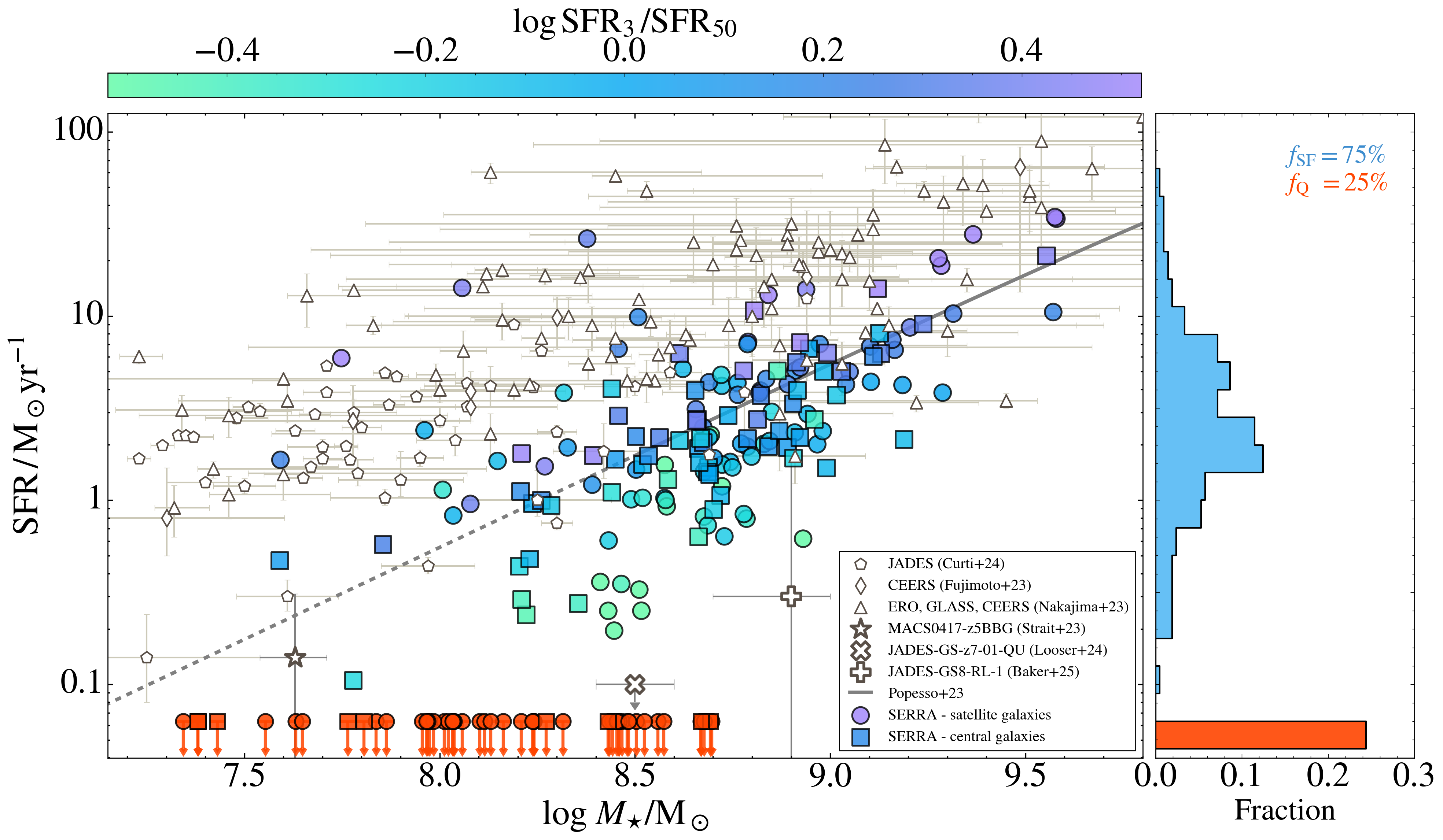}
\caption{Stellar mass ($M_\star$) - star formation rate (SFR), averaged over 3~Myr, relation for a sample of 208 galaxies at $z=6-8$ from the \code{SERRA} simulations \citep{Pallottini2022}.
The galaxies are color-coded with the ratio between SFR averaged over 3~Myr (SFR$_3$) and 50~Myr (SFR$_{50}$), which is an indicator of whether the galaxies are experiencing bursts ($>1$) or downturns ($<1$) of star formation. Galaxies undergoing phases of quiescence are shown in orange.
Different markers distinguish satellite galaxies orbiting more massive companions (squares, 131) from central galaxies (circles, 78).
We also show JWST results of $z\sim 4-9$ galaxies spectra from \cite{Curti23b, Nakajima23, Fujimoto23, Looser23, Strait23, Baker25} and the $z\sim 6$ star forming main sequence parametrization by \cite{Popesso23}, showing that JWST observations seem biased towards starbursting and high-SFR systems.
\label{fig:sfr_mstar}
}
\end{figure*}

\section{The physics of bursty galaxies}\label{sec:satellites_properties}

We here present the simulations' results for the sample of $6<z<8$, $M_\star<10^{9.5}\msun$ galaxies, analyzing their bursty evolution and its impact on the SFR$-M_\star$ relation and on the fraction of temporarily quiescent galaxies.

\subsection{Stellar mass build-up}

To understand how galaxies form and evolve in the simulations, in the main panel of Fig.~\ref{fig:cummass} we show the fraction of stellar mass formed as a function of the time elapsed from the first star formation event in the galaxy. The colors correspond to the final stellar mass $M_\star$ of each galaxy.
Low-mass systems ($M_\star<10^8\msun$) preferentially form all their stars in a few tens of $\Myr$, whereas the stellar mass build-up is much longer (up to $\sim 500~\Myr$) in more massive systems.

The rise in the mass of the galaxies is not always regular and smooth. Rather, most SFHs are characterized by both rapid rises due to starburst events and prolonged phases in which the galaxies are rendered \textit{temporarily quiescent} by feedback processes (horizontal tracks).

Most galaxies, even those with a final mass $M_\star > 10^9\msun$, experience periods of quiescence during the early stages of their evolution, typically within the first $\lesssim 200$~Myr of their lifetimes, when they are less massive, i.e. when $M_\star\lesssim10\%$ of their final stellar mass.
Lower-mass systems are in fact more sensitive to feedback and more easily quenched, being less efficient in retaining gas in their shallow potential wells after H$_2$ photoevaporation and SN explosions.

The time-variability of galaxies alternating between starbursting and quiescent phases can be quantified through their duty cycle, $f_{\rm duty}$, defined as the ratio of time spent in activity (SFR~$>0$) and the overall time since the formation of the galaxy (see Equation 1 in \citealt{Gelli23}). As shown in the inset, the average duty cycle increases with the stellar mass, indicating that it is more likely for lower-mass galaxies to undergo a period of temporary quiescence.

In the top panel, we show as an example the SFH of two typical galaxies of different masses, highlighting their bursty time-variable evolution. The higher-mass galaxy undergoes a phase of temporary quiescence after its first main burst of star formation. As it builds up its mass, its average star formation rate increases. While its evolution remains bursty, marked by peaks and downturns, the galaxy becomes more stable against feedback processes, preventing further quiescent phases. The lower-mass system is quiescent by the end of the simulation ($z\sim6$, marked by the dashed vertical line). Its previous evolution is similar to the first phases of the more massive galaxy, indicating that it may resume star formation again later on.

As we detect galaxies at specific times during their evolution, we want to study how their burstiness is expected to affect the observed SFR$-M_\star$ relation.

\subsection{SFR - stellar mass relation}\label{sec:mainseq}

In Fig.~\ref{fig:sfr_mstar} we show the SFR$-M_\star$ relation for the entire sample of  $M_\star \lesssim 10^{9.5} \msun$ galaxies at $z\sim6-8$.
Their SFR are in the range $\rm SFR = 0 - 50~\msunyr$, the average increasing with stellar mass, with the relation being characterized by a significant scatter, especially towards lower masses.
This scatter is the direct result of the bursty evolution that galaxies experience in their lifetimes, characterized by peaks of SFR (up to $\sim 20~\msunyr$ for galaxies with $M_\star \sim 10^8\msun$) and downturns in which the star formation is suppressed and can render some of them temporarily quiescent ($\rm SFR=0$, shown in orange).
To quantify the burstiness phase that each galaxy is experiencing, we color-code them according to the ratio between the instantaneous $\rm SFR_{3}$, averaged over the last 3~Myr, and $\rm SFR_{50}$, averaged over the last 50~Myr, similarly\footnote{While different definitions of burstiness can be adopted \citep{leja:2019, chavesmontero:2021, Pallottini23, Langeroodi24}, the physical interpretation of the result is qualitatively similar.} to \citet[][]{Endsley24}.
When ${\rm SFR}_{3}/{\rm SFR}_{50}>1$, the galaxy is undergoing a starbursting phase, while when the ratio is $<1$, the system is experiencing a downturn in star formation. 

At fixed stellar mass, galaxies with higher SFR values are typically in starburst phases, reaching $\sfrratio\sim 5$, whereas galaxies with lower SFR are more likely to be experiencing downturns in star formation.
Among these downturn galaxies, we note that only those with $M\star <10^{9}\msun$ experience strong suppression of star formation and can also undergo temporarily quiescent states.
This is a direct consequence of how burstiness affects galaxies at different masses, with low-mass galaxies more susceptible to feedback.

\begin{figure*}[t!]
\centering
\includegraphics[width=0.74\textwidth]{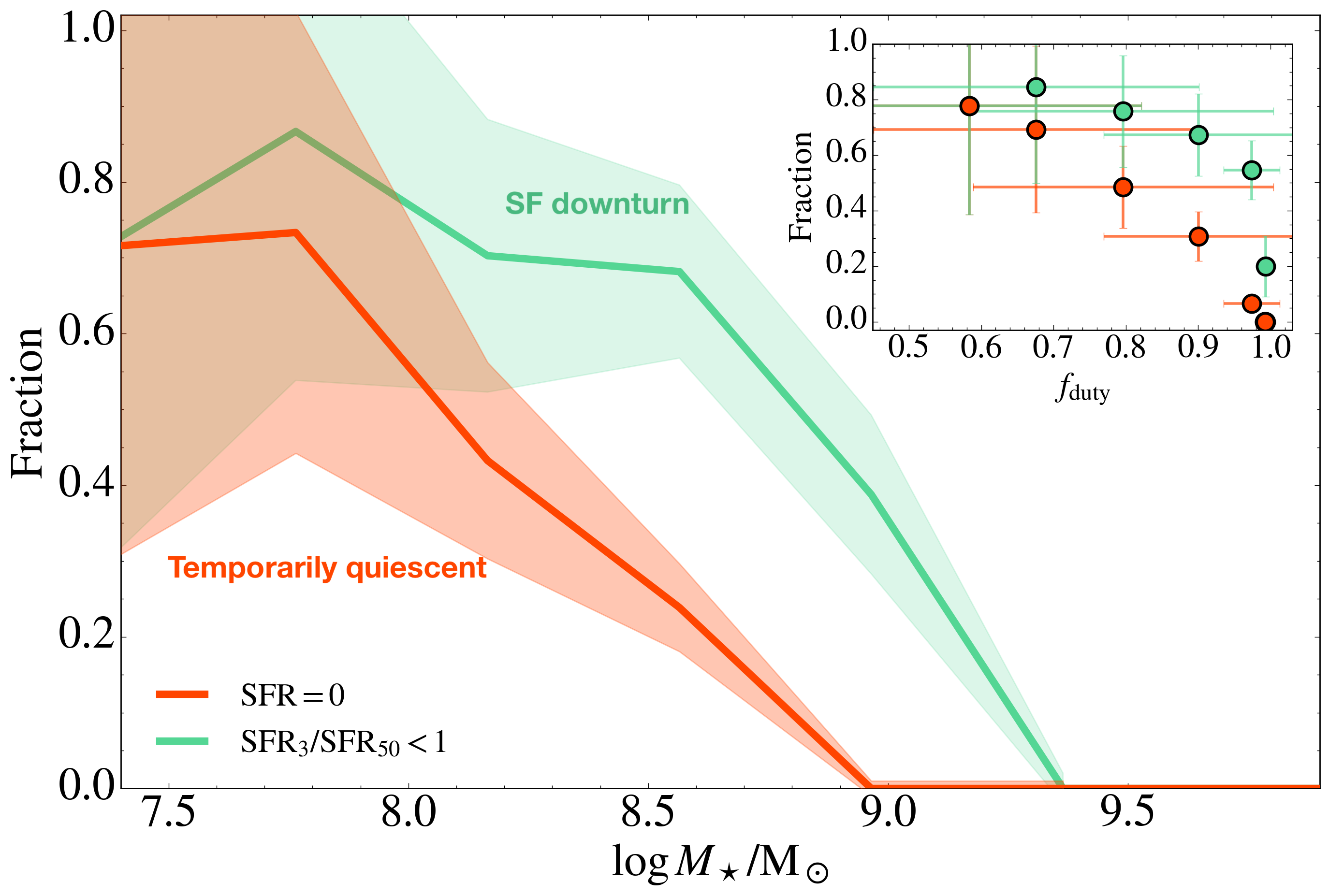}
\caption{Fraction of \code{SERRA} galaxies that are quiescent ($\rm SFR=0$) and that are experiencing a downturn in star formation ($\sfrratio<1$) galaxies per stellar mass bin. Shown are the 1$\sigma$ Poisson errors in each bin. The inset shows the same fractions as a function of the average duty cycle in each bin.
\label{fig:quiescentfraction}
}
\end{figure*}

The histogram in the right-hand panel displays the wide range of SFR covered by our sample of $M_\star = 10^7-10^{9.5}\msun$ galaxies.
Interestingly, we see that systems undergoing a temporarily quiescent phase constitute a significant fraction of $z\sim6-8$ galaxies, representing the $f_{\rm Q} = 25\%$ of the population.

To assess the potential impact of different environments on the bursty evolution of high-$z$ galaxies, in Fig.~\ref{fig:sfr_mstar} we distinguish central galaxies from satellite galaxies living in close proximity to a more massive galaxy.
Comparing their distribution in the $\mathrm{SFR}-M_\star$ relation, we find no large differences, with both types of galaxies roughly covering the same ranges of $M_\star$ and SFR.
In regions of high star formation (SFR $>10\ \mathrm{M_\odot\ yr^{-1}}$), satellites are more prevalent, but this is a statistical effect, as they constitute the majority of the sample (131 satellites to 78 central).
However, when comparing the fraction of temporarily quiescent galaxies in the two sub-samples we find that for satellites the fraction is $f_{\rm Q, sat} = 32\%$ and for central systems is $f_{\rm Q, isol} = 13\%$.
As shown by \citet{Gelli23}, both satellite and central galaxies undergo similar bursty feedback-regulated evolution, where the quenching is dominated by \textit{internal stellar feedback}. Indeed \citet{Gelli23} find that SNe feedback is the principal physical mechanism responsible for suppressing star formation in $M_\star < 10^{9.}\msun$ galaxies, typically acting on $\sim 30$~Myr timescales \citep[see also][in particular Fig. 4 therein]{Pallottini23}. This SN quenching is therefore independent of the environment in which they dwell.
The slight difference in the fractions of quiescent galaxies between the two samples is instead related to their ability to resume star formation after being quenched, which can depend on the environment. Specifically, we find that satellite galaxies are less likely to replenish their gas reservoirs, as most circumgalactic gas is accreted onto the nearby, more massive galaxy with its deeper gravitational potential.
Given the only minor differences between the two classes of galaxies, from now on we will not distinguish between the two and treat the sample of low-mass galaxies all together.

In Fig.~\ref{fig:sfr_mstar} we also show spectroscopic samples for JWST studies at similar redshifts \citep{Curti23b, Fujimoto23, Nakajima23}.
While the mass range covered by the observations and our simulations is comparable, the observed SFR range predominantly overlaps with the region of the highest star-forming simulated galaxies. Observations of low star forming galaxies ($\rm SFR\lesssim 1 \msunyr$) or undergoing downturns in star formation($\sfrratio\lesssim 1 \msunyr$) are rare.
This scarcity arises because time variability in the SFR causes fluctuations in galaxy luminosities, making galaxies in phases of declining star formation fainter and thus harder to detect (see Sec. \ref{sec:comparions_jwst}, particularly Fig. \ref{fig:endsley_comparison}).
As a result, observations, especially those relying on nebular line selection, are generally biased towards the highly star-forming tail of the galaxy population, and this intrinsic limitation should be carefully considered when inferring galaxy population properties in the EoR.

\begin{figure*}[t!]
\centering
\includegraphics[width=\textwidth]{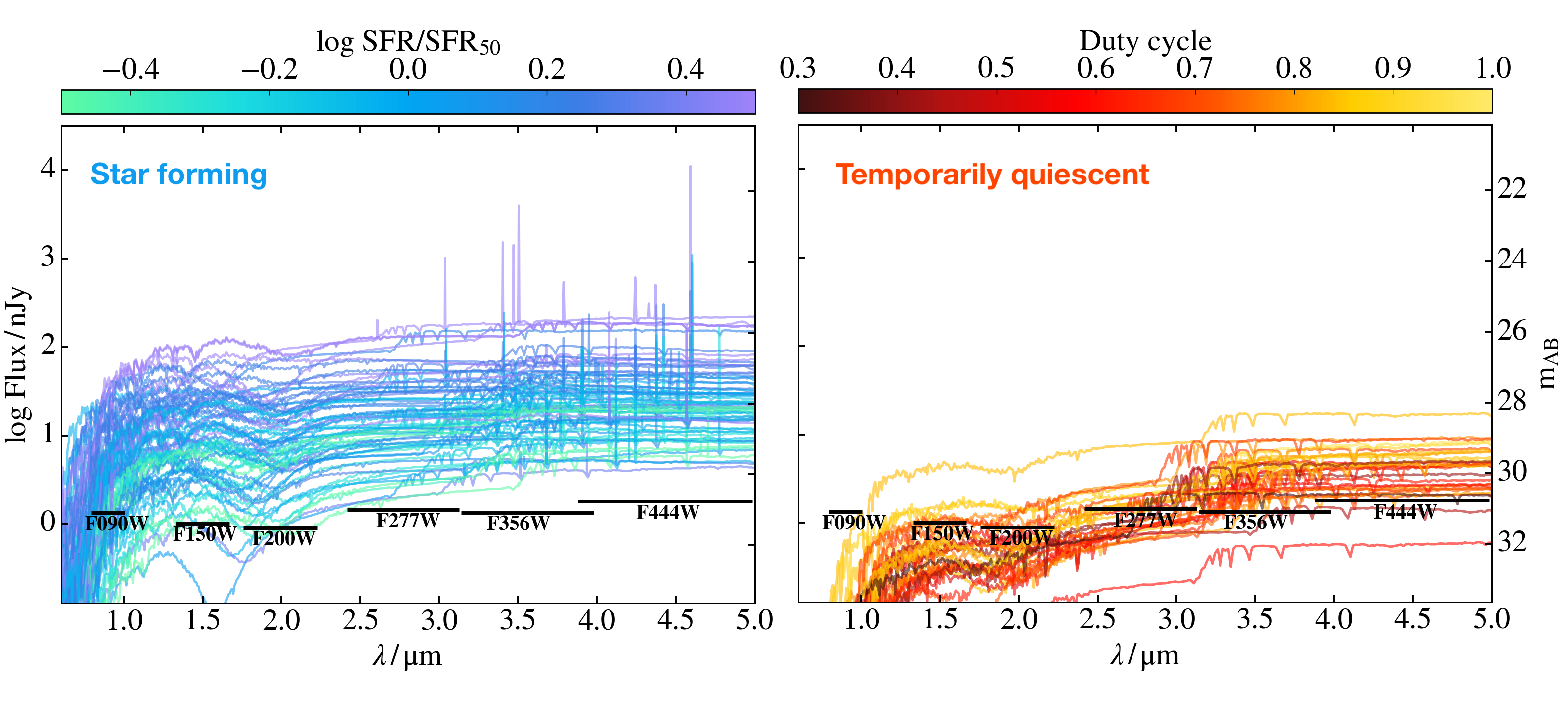}
\caption{Spectral energy distributions of the star forming and temporarily quiescent \code{SERRA} galaxies at redshift $6<z<8$ in the observed wavelength range covered by JWST/NIRCam and NIRSpec. Horizontal lines are the expected sensitivity limits for NIRCam imaging for a 20~hrs exposure observation with $\rm S/N\sim 3$. 
\label{fig:seds}
}
\end{figure*}

\subsection{Quiescent fractions}\label{sec:quiesc_frac}

The fraction of galaxies found in a quiescent phase at any given time is expected to be proportional to the overall duration of these periods, and inversely proportional to their duty cycle. Therefore, measuring the fraction of temporarily quiescent galaxies is crucial for estimating the duty cycles of high-$z$ galaxies, understanding the timescales of their star formation histories, and gaining insights into the physical processes that regulate them.

We study the mass dependency of the duty cycle of galaxies, by showing in Fig.~\ref{fig:quiescentfraction} the fraction of quiescent ($\mathrm{SFR}=0$) and downturn ($\sfrratio<1$) galaxies per stellar mass bin.
The fractions strongly increase with decreasing stellar mass. For $M_\star > 5\times 10^8 \msun$ quiescent systems represent $<10\%$ of the satellite population; at $M_\star \approx 10^{7.5} \msun$ the fraction becomes $70\%$. Similarly, the overall fraction of galaxies with recent SFR downturn follows a decreasing trend with the mass, going from $\gtrsim 80$\% at $M_\star \approx 10^{8} \msun$ to $\sim 30$\% at $M_\star \approx 10^9 \msun$.
The shaded regions show the $1\sigma$ Poisson error in each bin, which can reach unphysical values of $\gtrsim 100\%$ in the low-mass regime due to low number counts.

To highlight how these fractions are tightly linked to the bursty nature of the SFH, the inset shows the same fractions as a function of $f_{\rm duty}$ in each bin. At fixed stellar mass, higher fractions of quiescent galaxies reflect lower duty cycles, i.e. less time spent in activity.
This strong mass dependency once again reflects the fact that lower mass systems, typically residing in lower mass dark matter haloes, are especially sensitive to feedback processes, and their star formation can be more easily suppressed by stellar feedback through SN explosions and photoevaporation of H$_2$.

Measuring these fractions would provide a direct probe of high-$z$ bursty evolution, but remains highly challenging for several reasons. First, accurate measurements of stellar mass at high redshift are challenging due to the bursty nature of galaxies' SFH, which are difficult to capture with commonly used SED fitting codes \citep[e.g.][]{Markov23, Rusta24, Whitler2023}, as well as the outshining of older stellar populations by recent star formation \citep[e.g.][]{Narayanan24}.
Second, galaxies undergoing downturns or temporary quiescence are, on average, fainter than their star-forming counterparts of the same mass (see Sec.~\ref{sec:seds}) making it challenging to achieve the stellar mass completeness required to measure quiescent fractions, particularly at the low-mass end.

For these reasons, for a more reliable comparison between simulations and observations, we will compare fractions of quiescent galaxies at fixed luminosity rather than stellar mass.

In the following sections, we present the results of the forward modeling of the SEDs of \code{SERRA} galaxies, allowing us to predict the detectability of temporarily quiescent systems, compare with recent JWST observations, and explore the best observational strategies to better capture this elusive population.

\section{Constraining high-$z$ burstiness with JWST observations}\label{sec:seds}

Having analyzed the bursty evolution of galaxies that leads to the emergence of temporarily quiescent low-mass systems at $z > 6$, we now turn to investigating the impact of this evolution on their emission. In this section, we present the results of forward modeling the galaxies' emission, analyzing the SEDs of both star-forming and quiescent systems, and comparing them with observations.

\subsection{Bursty evolution shaping the SEDs}

In Fig.~\ref{fig:seds} we show the synthetic SEDs of the star-forming (left) and temporarily quiescent (right) galaxies (see Sec.~\ref{sec:sed_modelling} for the model).
The flux density is plotted as a function of the observed wavelength in the range $0.6 < \lambda/\mu m < 5 $, covered by the NIRCam and NIRSpec instruments on board JWST\footnote{The displayed SEDs show the emission without the inter-galactic medium absorption, that for galaxies at $z\sim6-8$ is expected to heavily affect the observed flux in the JWST/NIRCam F090W filter.}. 

The SEDs of star-forming galaxies, which are on average more massive than quiescent ones (Fig. \ref{fig:quiescentfraction}), are characterized by higher flux at all wavelengths and by the presence of prominent emission lines associated with ongoing star formation. Their SEDs are color-coded according to their $\sfrratio$, highlighting how galaxies undergoing suppressions/downturns in star formation tend to be less luminous and have weaker emission lines than those experiencing bursts.
By contrast, spectra of temporarily quiescent galaxies are not only fainter, but they also show strong Balmer breaks produced by the presence of older stellar populations and the lack of young stars contributing to the emission. They are characterized by redder colors, with the emission peaking at $\lambda \gsim 3~\mu m$. Among these temporarily quiescent systems, brighter galaxies are characterized by higher duty cycles, i.e. they have been only recently quenched and have spent the least time in the quiescent phase.

The effect of dust extinction is relevant in some of the star-forming systems with $M_\star \gtrsim 10^{8.5}\msun$, where column densities can reach $N_{\rm H} \sim 10^{22} \rm cm^{-2}$. Conversely, it is less important in quiescent galaxies where the amount of gas and dust is negligible having been mostly evacuated or heated by SN explosions.

It is now evident how galaxies experiencing a star-bursting phase are more luminous in all filters and easier to detect in large surveys, while only very deep observations can allow the detection of the less massive quiescent galaxies.
Typical NIRCam sensitivity limits for point sources with signal-to-noise ratio of $\rm S/N \sim 3$ and exposure time of $\sim 20~\rm hrs$ are also shown\footnote{Flux limits for a $10$~ks integration are retrieved from Table 1 of \url{https://jwst-docs.stsci.edu/jwst-near-infrared-camera/nircam-performance/nircam-sensitivity} (last updated on 25th November 2022), based on the Exposure Time Calculator (ETC) v2.0, assuming photometric apertures 2.5 pixels in radius and a benchmark background ($1.2\,\times$ minimum zodiacal light).}. The fact that many temporarily quiescent galaxies' SEDs are below these values, especially at wavelengths below $\lambda < 3 \mu m$, highlights how unveiling this population represents a challenge even for the deepest JWST surveys.

\subsection{Comparison with JWST observations}\label{sec:comparions_jwst}

\begin{figure}[t!]
\centering
\includegraphics[width=0.5\textwidth]{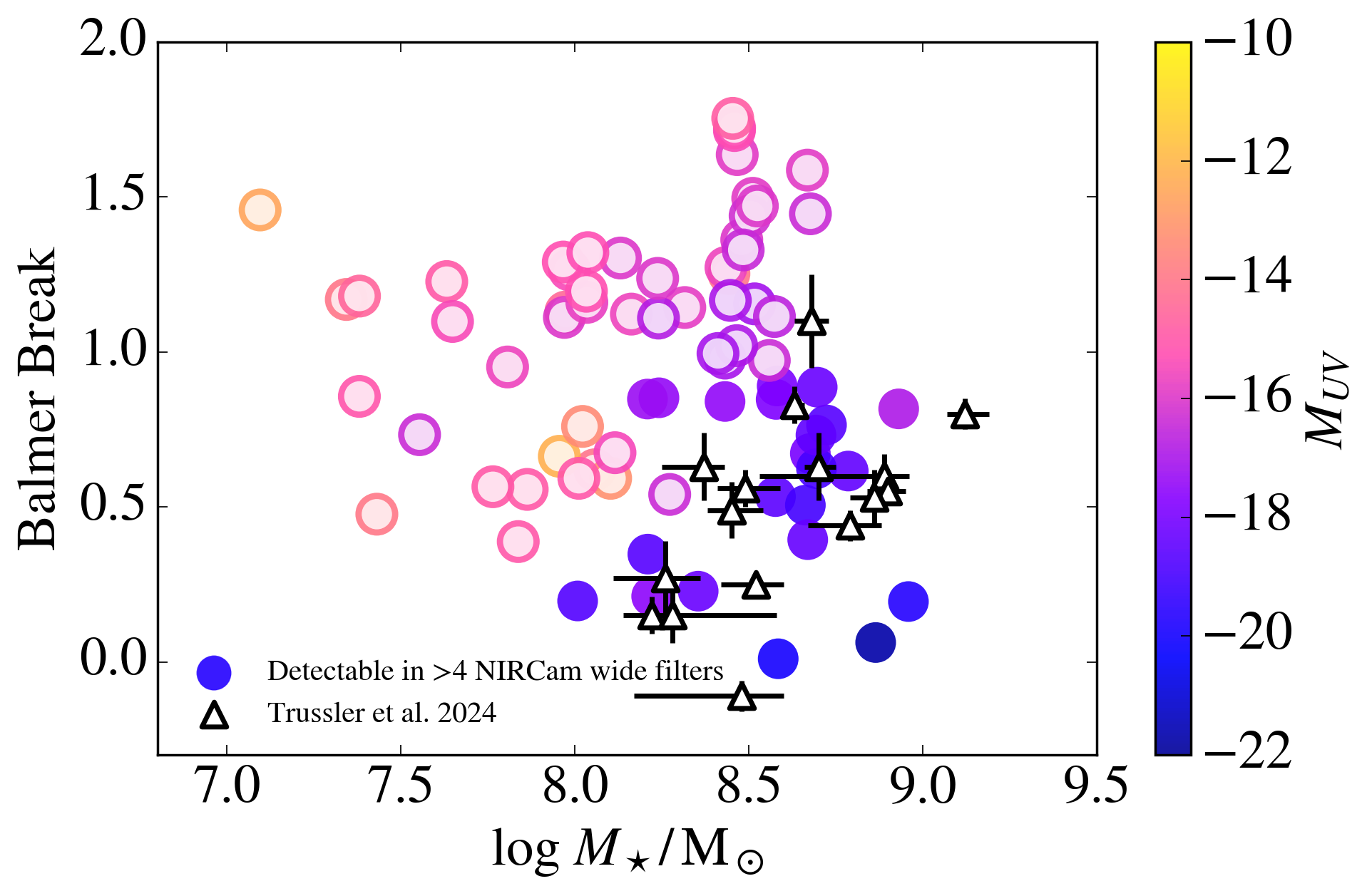}
\caption{Balmer break strength as a function of the stellar mass for temporarily quiescent galaxies in SERRA. The colors show the UV magnitude and filled scatter points mark those galaxies that are above typical deep (20~hrs) NIRCam detection limits in at least 4 wide band filters. Datapoints show observations from temporarily quiescent galaxies in JADES by \cite{Trussler24}.
\label{fig:balmerbreaks}
}
\end{figure}
We can now use the synthetic SEDs to translate our predictions for the fractions of temporarily quiescent and downturn galaxies analyzed in Sec.~\ref{sec:quiesc_frac} into observable properties.
In temporarily quiescent galaxy spectra, a key feature is a strong Balmer break, which strength we can define as the difference between the flux in the filters F277W and F444W.

In Fig.~\ref{fig:balmerbreaks} we show the strength of Balmer break as a function of the stellar mass, colored with the UV magnitude. The filled points identify the \code{SERRA} galaxies that would be above observational limits for $\sim$20~hrs imaging.
We compare with observations from \cite{Trussler24} that identified temporarily quiescent galaxies in JADES from narrow and medium-band imaging using their lack of emission lines. Interestingly, the observations cover the same Balmer break - $M_\star$ regions as our predicted \textit{detectable} temporarily quiescent galaxies, implying that detected galaxies are indeed sources undergoing drops of star formation due to burstiness.
Fig.~\ref{fig:balmerbreaks} also highlights how temporarily quiescent galaxies with similar stellar mass can exhibit a wide range of luminosities and Balmer breaks, and not all of them exceed the detection limits. For this reason, when comparing fractions of temporarily quiescent galaxies with observations, it is crucial to consider mass incompleteness. To address this issue, we can leverage our forward modeling to explore whether and how the stellar mass dependence of burstiness (see Fig.\ref{fig:quiescentfraction}) translates into a corresponding luminosity dependence \citep[e.g.][]{Sun24}.

\begin{figure*}[t!]
\centering
\includegraphics[width=0.95\textwidth]{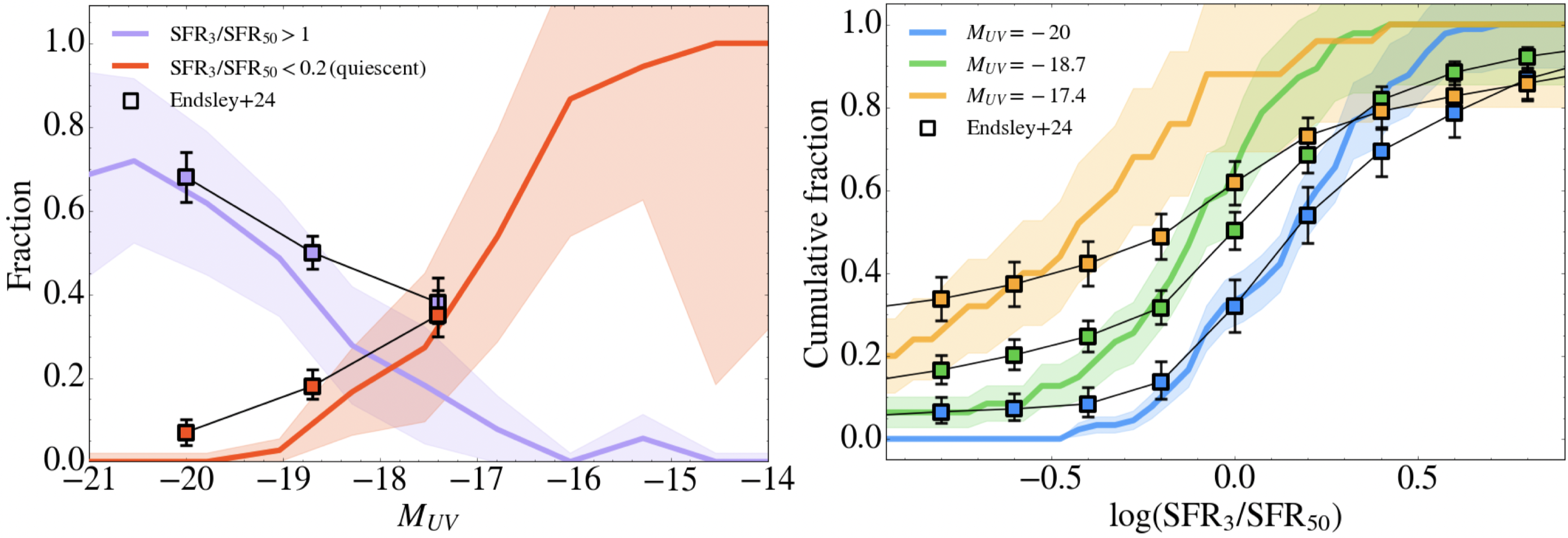}
\caption{Comparison between \code{serra} and observations by \cite{Endsley24}.
\textit{Left}: fractions of temporarily quiescent galaxies $\sfrratio < 0.2$ and starburst galaxies $\sfrratio >1$ as a function of the UV magnitude.
\textit{Right}: cumulative fractions of galaxies in three $M_{UV}$ bins as a function of the SFR ratio.
The trends confirm the mass dependence of bursty evolution, with low-mass, fainter galaxies more likely to experience quiescence. The slightly higher observed fraction of bursty galaxies suggests shorter duty cycles and enhanced burstiness in high-$z$ galaxies compared to the simulation.
\label{fig:endsley_comparison}
}
\end{figure*}

In a recent study, \cite{Endsley24} analyses the SFH of over 300 Lyman break galaxies at $z\sim 6$ down to $M_{UV}<-16$, finding a large diversity in $L_{H\alpha}/L_{UV}$, index of recent changes in SFR, among galaxies at the same magnitude.
In Fig.~\ref{fig:endsley_comparison} we compare our predictions for the fractions of bursty galaxies with \cite{Endsley24} observations. The left panel shows the fractions of galaxies with $\sfrratio > 1 $, i.e. that are experiencing recent bursts of star formation, and those with $\sfrratio < 0.2 $, i.e. that are experiencing extremely strong downturns and that in our sample correspond to temporarily quiescent galaxies.
\code{SERRA} predicts the fraction of quiescent galaxies to increase for faint galaxies, reflecting the stellar mass dependency (see Fig.~\ref{fig:quiescentfraction}), and to be the dominant ($>50\%$) population above $M_{UV}>-17$. On the other hand, galaxies with a recent increase in star formation are predicted to be more abundant at higher luminosities.

When comparing our predictions to the observations of \cite{Endsley24}, we find that the simulated trends for both types of galaxies align remarkably well with the observed fractions. However, while the increasing abundance of quiescent (starbursting) galaxies towards lower (higher) luminosities is well reproduced, their overall fraction is from 0.01 to 0.15 dex higher in observations compared to the simulation.

The fact that observed \quotes{extreme} galaxies, both starbursting and quiescent, are more abundant than the simulations indicates that high-$z$ galaxies have a more bursty evolution than predicted, with shorter duty cycles.

The right panel of Fig.~\ref{fig:endsley_comparison} shows the cumulative fractions of galaxies in three $M_{UV}$ bins as a function of their burstiness quantified through the ratio $\sfrratio$.
The fractions follow similar general trends, with fainter galaxies being dominated by downturn and quiescent population ($\sfrratio<1$), and brighter galaxies exhibit a steeper rise towards higher ratios. 
The observed trends, especially for the fainter galaxies, rise more slowly than the simulation, hinting once again to the somewhat stronger burstiness encountered in observations, causing higher fractions of galaxies to be populating the low $\sfrratio$ tail at all luminosities \footnote{We also note that part of the mismatch is to be attributed to the different overall sample analyzed in simulations and observations: while we selected galaxies with $M_\star<10^{9.5}\msun$ from \code{SERRA}, observed galaxies can reach higher masses and luminosities, the reason why the curves do not reach unity in the plot.}.

\begin{figure*}[t]
\centering
\includegraphics[width=\textwidth]{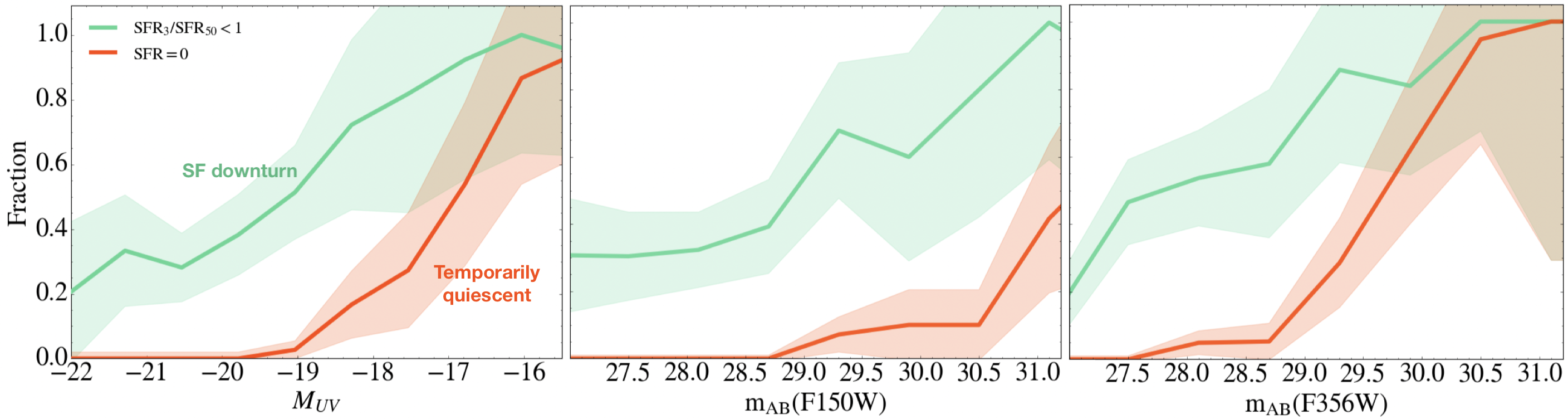}
\caption{Fractions of temporarily quiescent (red) and downturn (green) galaxies in the \code{serra} simulations as a function of the absolute magnitude $M_{UV}$ (left panel), and the AB magnitude in two NIRCam filters F150W (central panel) and F356W (right panel), respectively blueward and redward of the Balmer break.
\label{fig:fq_mag}
}
\end{figure*}

In Fig.~\ref{fig:fq_mag} we provide further predictions for the fractions of quiescent (SFR$ =0$) and downturn galaxies ($\sfrratio<1$) as a function of $M_{UV}$ and AB magnitudes in two JWST/NIRCam filters (F150W and F356W, blueward and redward of the Balmer break respectively).
As expected, the fraction of both low star forming and quiescent galaxies increases towards fainter sources. However, we notice that at fixed $\rm m_{AB}$, the fraction of quiescent galaxies detected in F356W is significantly larger than in F150W. This effect is due to the prominent Balmer break that characterizes these galaxies, making most of them appear only in the redder filters in imaging surveys.
In Section~\ref{sec:f200drop}, we explore how this may be exploited to unveil the faintest quiescent galaxies.

\section{Discussion}\label{sec:observationalstrategy}

\subsection{Interpretation of JWST observations}

We have seen how observations by \citet{Endsley23b} suggest a slightly burstier star formation in observed $z\sim6$ galaxies than what was predicted in \code{serra}.
Since the evolution and quenching of the simulated low-mass galaxies are largely dominated by internal stellar feedback from H$_2$ photoevaporation and SN explosions, the similar observed trends indicate that these are indeed the main factors driving bursty evolution, but that stronger feedback or additional processes may be at play to boost the burstiness and quenching of galaxies.

One possibility is that the additional required feedback is provided by the action of stronger radiation pressure from young stars \citep[as suggested in][]{Ferrara23, Ferrara24} also required to explain the observed spectra of the currently observed quiescent galaxies \citep{Looser23, Strait23}, as shown in \cite{Gelli23, Gelli24a}.

Moreover, \cite{Endsley23b} also reports the observation of galaxies undergoing extreme bursts of star formation with $\sfrratio>5$, which are not encountered in our sample. This discrepancy is likely primarily due to the stellar mass selection of the galaxies in our sample ($M_\star<10^{9.5}\msun$), but may also partially reflect a higher degree of burstiness in the observed galaxies.

Here, we have focused on galaxies in the final stages of the EoR, as this is the optimal regime where the fluxes of quiescent low-mass systems are high enough to be observed. However, the high level of burstiness inferred is also likely present at very high redshifts $z > 10$, where extreme star formation stochasticity is expected to play a role in enhancing the observed UV luminosities \citep[e.g.,][]{Gelli24b, Mason23, Pallottini23, Kravtsov2024}.

Our predictions also show how the fractions of temporarily quiescent galaxies are expected to increase even more at lower luminosities, reaching $~80\%$ at $M_{UV}\gtrsim -16$.
Pushing observations to fainter limits, e.g. with deeper observations or gravitational lensing, would therefore provide crucial insights into the feedback processes regulating bursty star formation. For instance, detecting higher fractions than predicted in this low-luminosity regime would suggest that SN feedback, acting in all galaxies after starbursts, needs to be stronger. Conversely, if the observed trends align with our predictions at $M_{UV} > -16$, it would indicate that the currently observed discrepancy is related to higher-mass galaxies, where different feedback mechanisms, such as AGN, may play a role.

Interestingly, JWST has also been revealing more massive quiescent galaxies at surprisingly high redshifts \citep[e.g.][]{Weibel24, Carnall23, deGraaff24, Kokorev24}. Explaining their nature is beyond the scope of this work as their quiescence is likely not temporary and not linked to bursty star formation, but rather to a stronger feedback mechanism such as AGN feedback. As black hole feedback is not included in \code{serra}, this interpretation would also be consistent with the fact that we do not find any massive quenched galaxy in our sample. In fact, if AGN feedback is indeed at play at $z>6$, we expect the fractions of quiescent galaxies shown in Fig.~\ref{fig:quiescentfraction} to increase once again in the high-mass regime, i.e., above $M_\star > 10^{10}\msun$.

\begin{figure*}[t!]
\centering
\includegraphics[width=\textwidth]{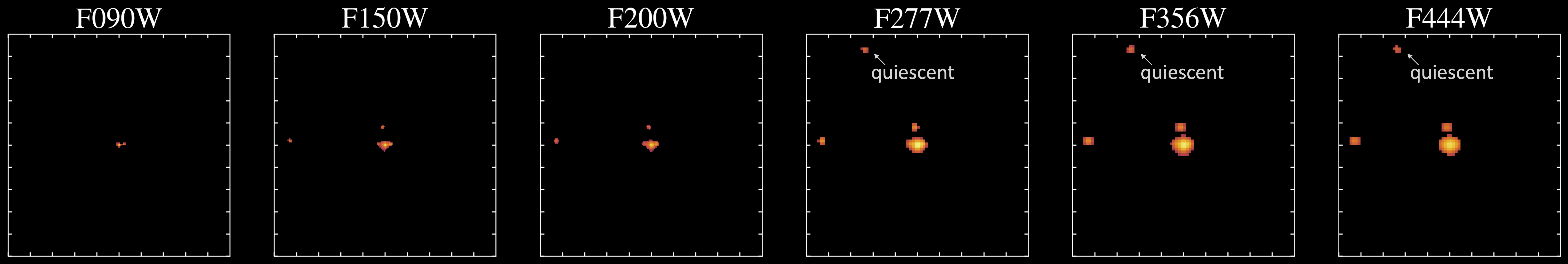}
\caption{Synthetic image of a group of galaxies at $z\sim6$ as they would appear in different JWST/NIRCam filters. The colored pixels are those with expected flux at $\rm S/N > 5$ in a typical 20~hrs observation. The displayed field of view has a side of $5\arcsec$ and is centered on a massive ($M_\star \sim 10^{10} \msun$) galaxies. One out of its three satellites is a temporarily quiescent F200W drop-out galaxy, as pinpointed by the white arrow.
\label{fig:image}
}
\end{figure*}
\subsection{Prospects: F200W drop-outs and satellites} \label{sec:f200drop}

Our predictions show that current observations are typically limited to the more luminous ($M_{UV}<-17$) among the temporarily quiescent galaxies, i.e. the ones that have only been recently quenched. This is expected to be the tip of the iceberg, as our analysis suggests the presence of a large population of hidden quiescent galaxies dominating at lower masses and fainter regimes ($M\star < 10^8\msun$, $M_{UV}>-17$).

To understand whether and how the faintest end of the population of temporarily quenched galaxies may appear in JWST observations, we produce mock NIRCam images for all our simulated systems.
In Fig.~\ref{fig:image} synthetic images of a $z\sim 6$ group of galaxies are shown in different NIRCam filters. Pixels are colored with increasing flux when the signal-to-noise exceeds $\rm S/N>5$ (black elsewhere), for an exposure time of 20~hrs. 

The image, of size  $5''\times 5''$, corresponding to $30~\kpc \times 30~\kpc$, is centered on a massive $M_\star \sim 10^{10}\msun$ star forming galaxy that is surrounded by three lower-mass satellites: two actively star-forming galaxies and one in temporary quiescence. This typical \highz dense environment allows us to understand how different sources may appear in deep JWST imaging surveys. 
All the low-mass bursty galaxies ($M_\star<10^{9.5} \msun$) appear as compact point sources, while the central more massive galaxy is more extended and resolved.

The temporarily quiescent galaxy, currently not forming stars and therefore dominated by old stellar populations, appears only in long wavelength filters, as a \textit{F200W drop-out} source. As typical temporarily quiescent SEDs (see Fig.~\ref{fig:seds}) are characterized by strong Balmer breaks, the observed flux drops drastically below $\sim3~\mu m$, where only upper limits can be placed even in $\sim 20$~hrs observation.
The vast majority of temporarily quiescent galaxies in our sample are predicted to appear as F200W drop-outs in typical deep surveys. In particular, mimicking a $\sim 20$~hrs exposure, we expect that out of 51 temporarily quenched galaxies, only 25 are bright enough to be observable in at least 3 NIRCam filters, and 21 of them, i.e. the 84\%, appear as F200W drop-outs.
This high fraction suggests that current observations may be overlooking the majority of quiescent galaxies, not identified with current selection methods. Instead, these currently primarily select the brightest among the quiescent galaxies, with emission spanning multiple filters and whose star formation was only recently suppressed.

Moreover, the predicted fraction of temporarily quiescent sources could be even larger if the star formation is more bursty than expected, as hinted by the results from \cite{Endsley24} and \citet{Ciesla23}.
Identifying F200W drop-out candidates in deep JWST fields may be the key to exploiting the full potential of current galaxy surveys and uncovering the hidden population of temporarily quiescent galaxies at high redshift.

Relying on photometry of F200W drop-out sources to identify temporarily quiescent galaxy candidates is a challenging task given that redshift estimates are inevitably characterized by large uncertainties, up to $\Delta z_{\rm phot}\sim 3$ (see App.~\ref{app:photoz} where we test the synthetic photometry with \code{bagpipes}).
However, this issue may be mitigated by exploiting high-$z$ galaxies \textit{overdensities}, like the one shown in the example in Fig.~\ref{fig:image}. We can in fact show that, despite the large photo-$z$ uncertainty, if a quiescent galaxy candidate is found at a close angular distance ($\lesssim 5\arcsec$) from a massive spectroscopically confirmed galaxy, then the probability of it being a satellite is as high as $\sim 95\%$ (see App.~\ref{app:photoz} for details), allowing to confirm its redshift as the same of the host massive galaxy.

As we have seen in Sec.~\ref{sec:mainseq}, the physical processes driving the quenching of low-mass galaxies are the same for satellite and central dwarf galaxies (i.e. internal feedback processes). However, low-mass quenched galaxies in such dense environments may more likely experience \quotes{starvation}, as most of the fresh gas from the circumgalactic medium is funneled toward their more massive companions. Consequently, the environment may affect the ability of quenched galaxies to resume star formation and undergo \quotes{rejuvenation} \citep[e.g.][]{Witten25}, a scenario we predict to be more likely for central galaxies.
Thus, detecting candidates in $z>6$ galaxy overdensities would allow us to both robustly identify a population of quiescent galaxies, and assess potential environmental effects.

\section{Summary and conclusions}\label{sec:summary}

Using the \code{SERRA} suite of cosmological simulations we have shown that, as a result of a highly bursty star formation, \textit{temporarily quiescent} low-mass galaxies are expected to be common in the Epoch of Reionization. We built a sample of $>200$ galaxies ($M_\star < 10^{9.5}M_\odot$), between $6\lesssim z \lesssim 8$; by analyzing their physical and evolutionary properties we find that:

\begin{itemize}

\item bursty star formation leads to a large scatter in the $SFR-M_\star$ relation, increasing toward lower masses, with $\sigma_{\log \rm SFR/\langle SFR \rangle} \sim 0.4$ for $M_\star \sim 10^8 \msun$;

\item low stellar-mass galaxies are easily quenched due to their shallow gravitational potentials, and the fraction of temporarily quiescent galaxies increases with decreasing mass, reaching $ \gtrsim 50\%$ of the population for $M_\star<10^8 \msun$;

\item both \textit{central} and \textit{satellite} galaxies experience temporary quiescence due to internal feedback, indicating that the environment does not play a key role in their quenching. However, the higher fraction of quiescent satellites (32\%) compared to centrals (13\%) suggests that nearby massive companions may inhibit the efficient re-accretion of fresh gas favoring starvation (rejuvenation) in satellite (central) galaxies;

\end{itemize}

We derived the synthetic SEDs of the simulated galaxies taking into account both continuum and line emission, as attenuated by dust, to compare with JWST observations and make future predictions. 
The main results are: 

\begin{itemize}

\item temporary quiescent galaxies are typically faint ($\langle M_{UV}\rangle = -15.6$ for $M_\star \sim 10^8\msun$) with a broad range of UV luminosities at fixed stellar mass, and their spectra lack emission lines but exhibit strong Balmer breaks (0–2 for $M_\star \sim 10^{8.5}\msun$), consistent with observations by \cite{Trussler24};

\item the fraction of temporarily quiescent (starburst) galaxies rapidly increases (decreases) toward faint luminosities, becoming the dominant population above $M_{UV} > -17$;

\item predicted fractions for temporarily quiescent galaxies follow similar trends to those observed by \cite{Endsley24}, but are slightly lower (between 1 and 15 \%) at $-20<M_{UV}<-19$. The high abundance of extremely starbursting and quiescent galaxies observed means that $z\sim 6$ galaxies undergo an even burstier star formation than in our simulations, suggesting that on top of the dominant SNe, stronger feedback or additional processes may be at play \citep[e.g.,][]{Gelli23};

\item most temporarily quiescent galaxies at $z\sim 6-8$ are predicted to be detectable as F200W drop-outs in current deep surveys, with galaxy overdensities offering promising environments to locate them as satellites of massive sources, enabling high-confidence photometric identification ($95\%$).

\end{itemize}

Our results convincingly show that stellar feedback plays a vital role in shaping the SFHs of galaxies during the Epoch of Reionization.
Bursty star formation significantly influences the observability of early galaxies over time, introducing a bias in our view of the early Universe toward the more luminous starbursting sources at high redshifts. Accurate predictions of these effects are crucial to be able to effectively interpret observations of galaxy populations at the highest redshifts. 
JWST has begun to reveal the tip of the iceberg of the vast population of faint less star-forming galaxies at the edge of reionization. As our models predict these galaxies to be the majority at lower luminosities, upcoming unprecedentedly deep surveys \citep[e.g. GLIMPSE][]{glimpse_jwst} and specific strategies to target temporary quiescent galaxies will be fundamental to unveil this extremely faint population and expand our knowledge on early galaxy evolution.

\section*{Acknowledgements}
We thank Ryan Endsley for sharing his data.
SS acknowledges support by the ERC Starting Grant NEFERTITI H2020/804240 (PI: Salvadori).
VG and CAM acknowledge support by the Carlsberg Foundation under grant CF22-1322. The Cosmic Dawn Center (DAWN) is funded by the Danish National Research Foundation under grant DNRF140.
This work is supported (AF) by the ERC Advanced Grant INTERSTELLAR H2020/740120,
and in part by grant NSF PHY-2309135 to the Kavli Institute for Theoretical Physics (KITP).
%
%
%
We acknowledge the CINECA award under the ISCRA initiative, for the availability of high-performance computing resources and support from the Class B project SERRA HP10BPUZ8F (PI: Pallottini).
We gratefully acknowledge the computational resources of the Center for High Performance Computing (CHPC) at SNS.
We acknowledge usage of the Python programming language \citep{python2,python3}, Astropy \citep{astropy}, Cython \citep{cython}, Matplotlib \citep{Hunter2007}, NumPy \citep{VanDerWalt2011}, \code{pynbody} \citep{pynbody}, and SciPy \citep{scipy}.

\bibliography{refer,codes}
\bibliographystyle{aasjournal}

\appendix \label{app:photoz}

\begin{figure}
\centering
\includegraphics[width=0.35\textwidth]{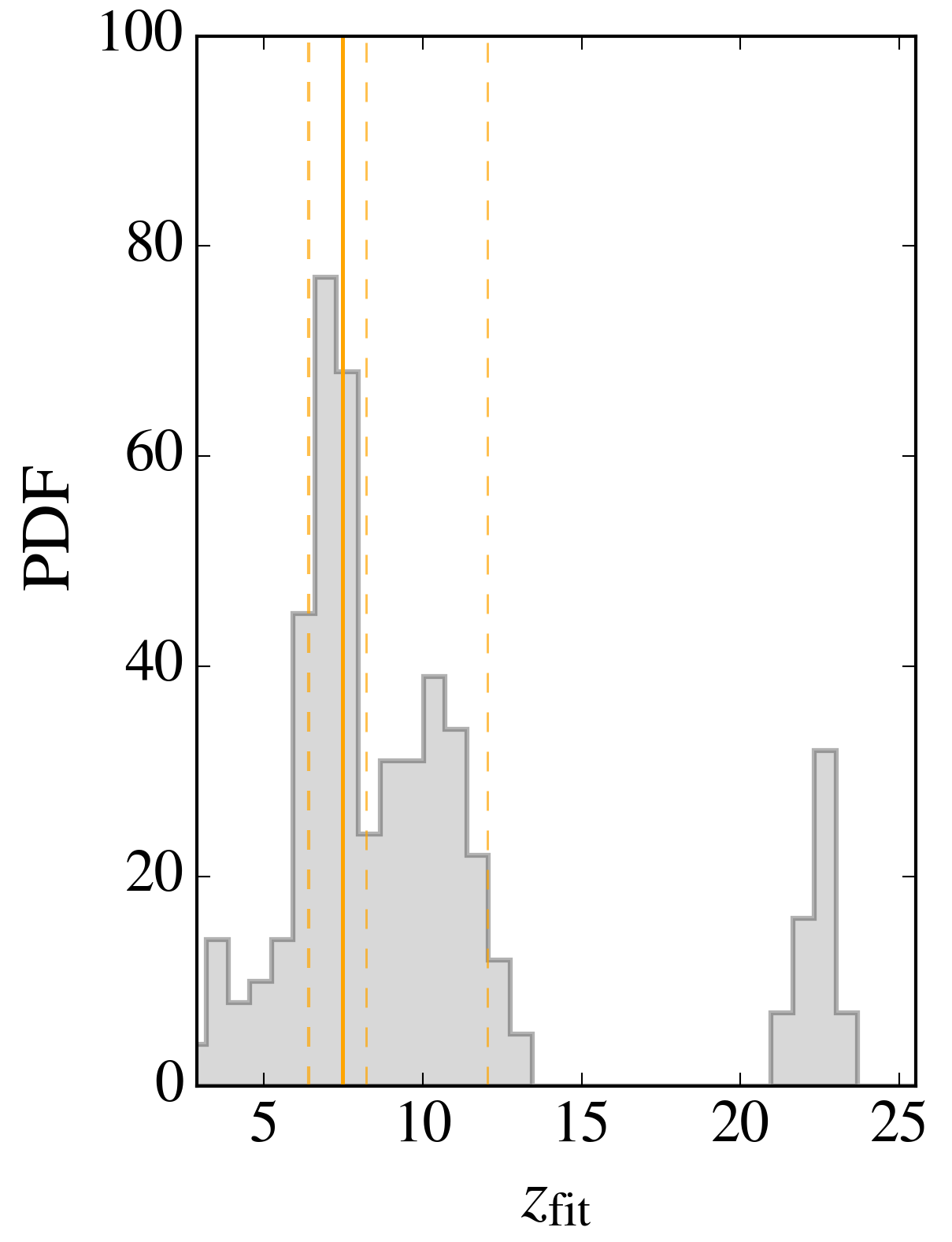}
\caption{\code{BAGPIPES} results for the photometric redshift for a typical temporarily quiescent galaxy.
\label{fig:bagpipesresults}
}
\end{figure}

Based on our results, a promising strategy to find temporarily quiescent galaxies is based on broad band photometry to identify F200W drop-out sources. First, we test the expected typical photo-z uncertainties by performing \code{bagpipes} \citep{carnall_bagpipes} SED fitting to mock photometry of our quiescent sources. Fig.~\ref{fig:bagpipesresults} shows the resulting probability distribution for the redshift of a typical F200W drop-out at $z=7$. The distribution is peaked at the correct redshift but is very broad leading to large uncertainties of $\Delta z_{phot} \sim 3$.

However, relying on dense environments in close proximity ($<5\arcsec$) of spectroscopically confirmed massive galaxies would put stronger constraints on the actual redshift of the F200W drop-outs. In fact, confirming that a photo-detected source is a satellite of a known massive galaxies would confirm its redshift as the same as the host galaxy's.
We need to address the question: what is the probability of a source in the proximity of a massive high-$z$ galaxy to be a satellite galaxy and not a foreground or background interloper?

\begin{figure}
\centering
\includegraphics[width=0.5\textwidth]{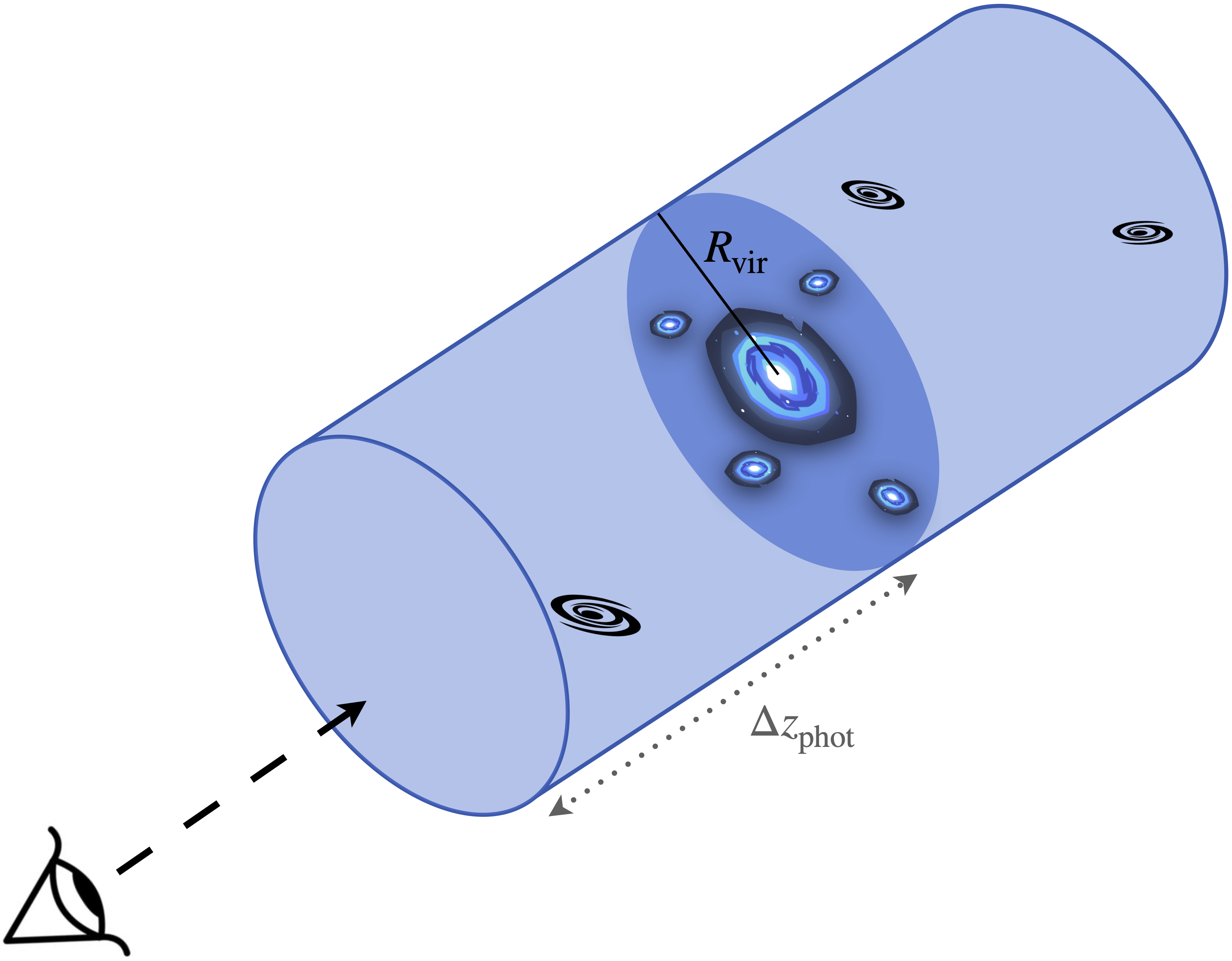}
\caption{Simple sketch illustrating the possible presence of background or foreground contaminant interloper galaxies (black spiral symbols) in the cosmic volume enclosed by the same redshift range measured for the satellites in the observed region around an LBG.
\label{fig:zphot}
}
\end{figure}

We can give an answer to this question using a simple probabilistic argument.
Given a certain photometric redshift error $\Delta z_{phot}$ compatible with the one of a target high-$z$ LBG, we want to evaluate the probability of a certain source detected in its vicinity (within $< R_{vir}$) to be one of its satellites and not an interloper field galaxy within the same redshift range. This can be quantified as $P(sat) = \frac{\langle N_{sat} \rangle}{\langle N_{sat} \rangle + N_{field}}$, where $ \langle N_{sat} \rangle$ is the average number of expected detectable\footnote{We assume a typical deep observation of 20~hrs reached in e.g. JADES as done in Sec.~\ref{sec:seds}} satellites per LBG system, and  $N_{field}$ is the number of the expected detectable field galaxies in within the same redshift range. To evaluate the latter we consider the number of galaxies at redshift $z$ in a given magnitude range and cosmic volume $V$ as\\
\begin{equation}
N_{field} = \int_{M_{min}}^{M_{max}} \phi_z(M) \, dM   \times V,
\end{equation}
where $\phi_z(M)$ is the Schecter luminosity function at redshift $z$ in units $\rm mag^{-1} Mpc^{-3}$ \citep[][]{Bouwens21}. As $M_{min}$ and $M_{max}$ we consider, respectively, the magnitude of the most luminous of our simulated satellites, and the magnitude corresponding to the typical expected sensitivity limit of JWST/NIRCam.
The redshift uncertainty enters through the considered cosmic volume $V$ (see the simple illustrative sketch in Fig.~\ref{fig:zphot}), given by the two-dimensional area of the field-of-view encompassed within the virial radius, times a depth along the line-of-sight due to the photometric redshift expected error: $V = \pi R_{vir}^2 (D_L(z + \Delta z_{phot}/2) - D_L(z - \Delta z_{phot}/2))$, where $\Delta z_{phot}$ is taken from our \code{bagpipes} estimates.

Considering the ``worst case scenario" (i.e. maximizing the possible number of interlopers) with: maximum virial radius ($R_{vir}=25$~kpc), the detectability limit of the deepest JWST surveys in the wide filter F200W at S/N=3, $\Delta z_{phot}$ as derived by using NIRCam broad bands only, we get $N_{sat} = 2.5$ and $N_{field} = 0.13$. We, therefore, obtain that the probability of a source with compatible photometric redshift and located near a high-$z$ LBG in the plane of the sky to be a satellite dwarf galaxy is $95 \%$. This confirms that the one proposed is a solid method to identify satellite candidates and infer their properties. Moreover, this argument should be kept in mind especially when identifying passive dwarf satellites: given their red colors they could be easily mistaken for higher redshift ($z>15$) galaxies. Hence, particular attention needs to be paid to such systems located at small angular distances from LBGs.

\end{document}